\documentclass[twocolumn,showpacs,preprintnumbers,amsmath,amssymb,prl]{revtex4}
\usepackage{graphicx}% Include figure files
\usepackage{dcolumn}% Align table columns on decimal point
\usepackage{bm}% bold math

\begin{document}

\preprint{APS/123-QED}

\title{Magnetic phases of a highly frustrated magnet, ZnCr$_2$O$_4$, up to 

an ultra-high magnetic field of 600 T}% Force line breaks with \\

\author{Atsuhiko Miyata}
\author{Hiroaki Ueda}%
 \altaffiliation[Present address: ]{Department of Chemistry, Graduate School of Science, Kyoto University, Kyoto, 606-8502, Japan}
 
\author{Yutaka Ueda}%
 
\author{Hironobu Sawabe}%
 
\author{Shojiro Takeyama}
 \email{takeyama@issp.u-tokyo.ac.jp}
 \affiliation{Institute for Solid State Physics, University of Tokyo, 5-1-5, Kashiwanoha, Kashiwa, Chiba, 277-8581, Japan}

%\affiliation{

\date{\today}% It is always \today, today,

\begin{abstract}
The Faraday rotation and magneto-optical absorption spectral measurements were conducted to reveal the full-magnetization process and map out a magnetic phase diagram of a typical geometrical frustrated magnet, ZnCr$_2$O$_4$, by using the electromagnetic flux compression method in ultra-magnetic fields up to 600 T. A fully polarized ferromagnetic phase is observed in which the absorption spectra associated with an exciton-magnon-phonon transition disappears. Furthermore, prior to the fully polarized ferromagnetic phase above 410 T, we found a novel magnetic phase above 350 T followed by a canted 3:1 phase. 

\end{abstract}

\maketitle

%\section{\label{sec:level1}First-level heading:\protect\\ The line
%break was forced \lowercase{via} \textbackslash\textbackslash}
Geometrically frustrated magnet systems have been a fascinating subject for decades because they have macroscopically degenerated ground states 
vulnerable to tiny perturbations (spin-lattice interaction, quantum and thermal fluctuation, etc.), which leads to the emergence of exotic magnetic 
states~\cite{moessner02, chubukov91}. Interestingly, magnetic fields also cause the appearance of peculiar magnetic phenomena such as the 
magnetization plateau~\cite{kageyama99, okamoto11}, 
magnetic supersolid~\cite{miyata11} and crystalline magnon state~\cite{tsunetsugu04} due to competition between a magnetic field and frustrated spin 
exchange interaction. In 
order to understand the essence of the geometrical frustration, it is indispensable to conduct magnetization measurements up to full magnetization. 
Most geometrically frustrated magnets show $T_\text{N}/|\theta_\text{CW}| << 1$, where $T_\text{N}$ and $\theta_\text{CW}$ are the Neel temperature and 
the Curie-Weiss temperature, respectively, 
because geometrical frustration suppresses the occurrence of a magnetic order. This inequality is the main reason that investigating the 
full-magnetization processes of geometrically frustrated magnets has required multiple extreme conditions, such as a high magnetic field and 
extremely low temperature~\cite{kageyama99, okamoto11}. Here, we developed magnetization measurements by the Faraday rotation method and 
magneto-optical absorption 
spectral measurements in photo-wavelengths in the visible light range in magnetic fields up to 600 T and at temperatures down to 5 K by the electro-magnetic flux compression (EMFC) 
method~\cite{takeyama10}. Geometrically frustrated magnets mainly consist of transition metal ions, and d-d transitions and the transitions 
associated with them, 
such as exciton-magnon transitions, are sensitive to the transformation of crystal and magnetic structures. In addition, most of these transitions 
can be observed in wavelengths in the visible light range. We applied the magneto-optical methods to a typical three-dimensional geometrically frustrated magnet 
with a pyrochlore lattice, ZnCr$_2$O$_4$ ($T_\text{N}=12.5$ K and $\theta_\text{CW}=-390$ K)~\cite{lee00, lee02}, and revealed the full-magnetization 
processes at different temperatures.

Cr$^{3+}$ ions in a chromium spinel oxide, $A$Cr$_2$O$_4$ ($A$=Zn, Cd, Hg), reside on a pyrochlore lattice, and therefore these compounds are 
regarded as a 
candidate for the three-dimensional frustrated magnet. Magnetization measurements in these compounds have revealed various magnetic phases, including 
a robust 1/2 plateau phase~\cite{miyata11,ueda05, ueda06, kojima08}. Penc {\it et al.} suggested that a spin-lattice interaction plays a key role in 
determining the magnetic phases 
under magnetic fields in these compounds~\cite{penc04}. 
The observed magnetization data of HgCr$_2$O$_4$ and CdCr$_2$O$_4$ showed a fairly good agreement with those of the theory with taking account of a 
large spin lattice coupling parameter, i.e., $b\approx0.1$~\cite{ueda06, kojima08, motome06}. Whereas, when $b$ is small which is the case for 
ZnCr$_2$O$_4$ ($b\approx0.02$), the theory predicted the realization of new phase, the canted 2:1:1 phase. This is what we have actually 
found at a lower magnetic field side of the 1/2 plateau phase~\cite{miyata11}.
ZnCr$_2$O$_4$ approaches 
an ideal frustrated 
magnet with only a nearest neighbor exchange interaction, compared with other sister compounds. The magnetization processes of an ideal frustrated 
pyrochlore antiferromagnet ($b=0$) have been unclear both experimentally and theoretically. In addition, S. Kimura {\it et al.} recently reported that 
another phase might exist between a canted 3:1 phase and the ferromagnetic phase in HgCr$_2$O$_4$, and the magnetic phases would be more complicated than 
those by the theory with a spin-lattice coupling~\cite{kimura11}. Therefore, full-magnetization measurements of ZnCr$_2$O$_4$ are needed.

Recently, we developed a copper-lined coil for the EMFC method, which enabled us to generate a magnetic field of over 
700 T~\cite{takeyama10}. 
We applied this system to measurements of solid-state physics combined with a low-temperature apparatus. We employed a hand-made liquid-He flow-type 
miniature cryostat made completely of ``stycast'' resin, and achieved a temperature down to 5 K~\cite{miyata11}. The sample temperature was monitored 
by an Au-Fe/Chromel thermocouple attached adjacent to the sample, but the value could be higher than the actual sample temperature by about 1 -- 2 K 
due to the temperature gradient between the sample and the thermocouple. The values of magnetic fields were measured by a calibrated pick-up coil 
wound at the vicinity of the sample. The estimated error of the magnetic field was about 3$\%$. A single crystal of ZnCr$_2$O$_4$ was grown by a 
vapor transport technique. The sample was cut parallel to the (111) crystal surface, attached on a quartz substrate and polished to a thickness of 
about 100 $\mu$m for the Faraday rotation measurement or 50 $\mu$m for  optical transmission measurements, and the diameter of the sample was about 1.5 mm. 
The Faraday rotation measurement was performed in a manner similar to the measurement described in Ref.~\cite{miyata11}. A semiconductor laser (a Coherent ``Cube'') of 635 nm wavelength was used as a light source. The incident light was linearly polarized to the horizontal axis. The transmitted light was separated into vertical ($I_\text{s}$) and horizontal ($I_\text{p}$) components by a Wollaston prism. The Faraday rotation angle of the sample ($\theta_\text{M}$) was determined by subtraction of the diamagnetic contribution from the sample and a quartz substrate from the raw data of $\theta_\text{F} = \frac{1}{2} \cos^{-1} \left(\frac{I_\text{p}-I_\text{s}}{I_\text{p}+I_\text{s}}\right)$. The magneto-optical absorption spectra measurements were performed using a streak camera system coupled with a CCD detector and a polychromater. A Xe arc-flash lamp was used as a light source and light was guided into and out of the sample by optical fibers as shown in Ref.~\cite{miura98}.

\begin{figure}
   \includegraphics[width=.77\columnwidth]{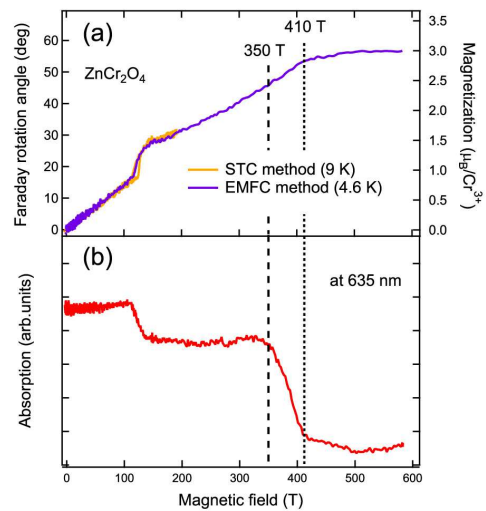}% Here is how to import EPS art
\caption{\label{fig:magnetization4k}(a) Magnetization obtained by the Faraday rotation method in magnetic fields of up to 600 T generated by the EMFC method at 4.6 K and up to 190 T by the STC method at 9 K as described in Ref.~\cite{miyata11}. (b) Absorption intensity change at a wavelength of 635 nm at 4.6 K as a function of the magnetic field.  The dotted line at 410 T indicates the point of phase transition into a ferromagnetic phase. The dashed line at 350 T indicates the onset of decrease of absorption intensity.
}
\end{figure}

In Fig.~\ref{fig:magnetization4k}, we obtained the Faraday rotation angle of the sample, which showed good agreement with that by the single-turn 
coil (STC) method up to 190 T.  We could obtain the magneto-optical absorption derived from the intensity of the monochromatic transmitted light in 
our experimental setup while measuring the Faraday rotation. In a magnetization curve, we clearly observed the transition into a ferromagnetic phase 
with a  saturated magnetization moment of 3 $\mu_\text{B}/\text{Cr}^{3+}$ at 410 T, and no clear anomaly was observed between 410 T and 160 T above 
the transition field from a 1/2 plateau phase to a canted 3:1 phase~\cite{miyata11}. The slope of magnetization in the antiferromagnetic phase 
between 0 T and 
120 T was larger than that in the canted 3:1 phase above 160 T. This is a point of difference from that in the theory taking into account the 
spin-lattice coupling by K. Penc {\it et al.} and D. L. Bergman {\it et al.}~\cite{penc04, bergman06}. This discrepancy implies that the exchange 
interactions should change due 
to lattice distortion associated with a magnetic phase transition, which agrees with the idea that the 1/2 plateau phase exists widely even above the Neel 
temperature ($T_\text{N} =12.5$ K), as described in Ref.~\cite{miyata11}.

On the other hand, we observed a rapid decrease of absorption between 350 T and 410 T. The absorption peak corresponding to a laser wavelength of 
635 nm arises from both a d-d transition ($^4A_2 \rightarrow ^4T_2$) and an exciton-magnon-phonon (EMP) transition in Cr$^{3+}$~\cite{szy80}. The 
energy and spectrum shape of these transitions are sensitive to the crystal and magnetic structures. Apparently, the anomaly at 120 T of the 
absorption is related to lattice distortion caused by a first-order transition from the antiferromagnetic phase to the canted 2:1:1 phase. Likewise, 
the rapid decrease of absorption between 350 T and 410 T implies a change in the crystal or magnetic structure. According to our recent work, there 
exists a canted 3:1 phase above 160 T~\cite{miyata11}, and the theories of K. Penc {\it et al.} and D. L. Bergman {\it et al.} allow no magnetic 
phases other than the ferromagnetic phase to exist in magnetic fields higher than a canted 3:1 phase~\cite{penc04, bergman06}. Therefore, the present 
data are indicative of the certain existence of a novel phase which has not yet been predicted by the conventional theories. To clarify this 
behavior, we investigated the temperature dependence of the magnetization and magneto-optical absorption measurements.

\begin{figure}
   \includegraphics[width=.75\columnwidth]{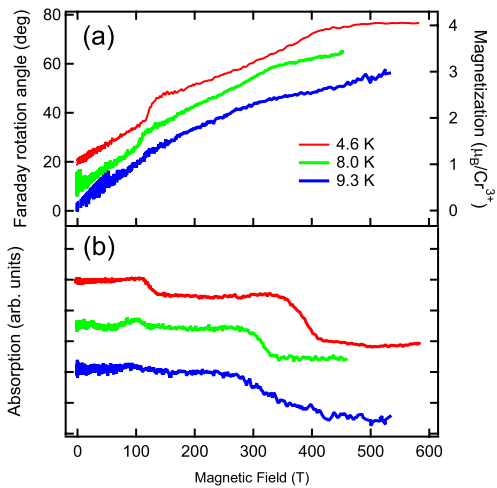}% Here is how to import EPS art
\caption{\label{fig:temp_mag} (a) Magnetization processes obtained from the Faraday rotation by the EMFC method measured at several temperatures. (b) Absorption intensity change in the wavelength of 635 nm at several temperatures as a function of magnetic field. }
\end{figure}

  Figure ~\ref{fig:temp_mag} shows the magnetization curve and magneto-optical absorption change in 635 nm at 4.6 K, 8.0 K and 9.3 K, respectively. 
In the magnetization curve, we observed the transition to the ferromagnetic phase at 330 T at 8.0 K, and no clear anomaly was observed at 9.3 K. This 
transition point shifted toward a lower magnetic field with increasing temperature. In addition, a decrease of the absorption at 295 T at 8.0 K and 
at 270 T at 9.3 K were observed, and the field at which the absorption decreased shifted lower as the temperature increased.

\begin{figure}
   \includegraphics[width=.68\columnwidth]{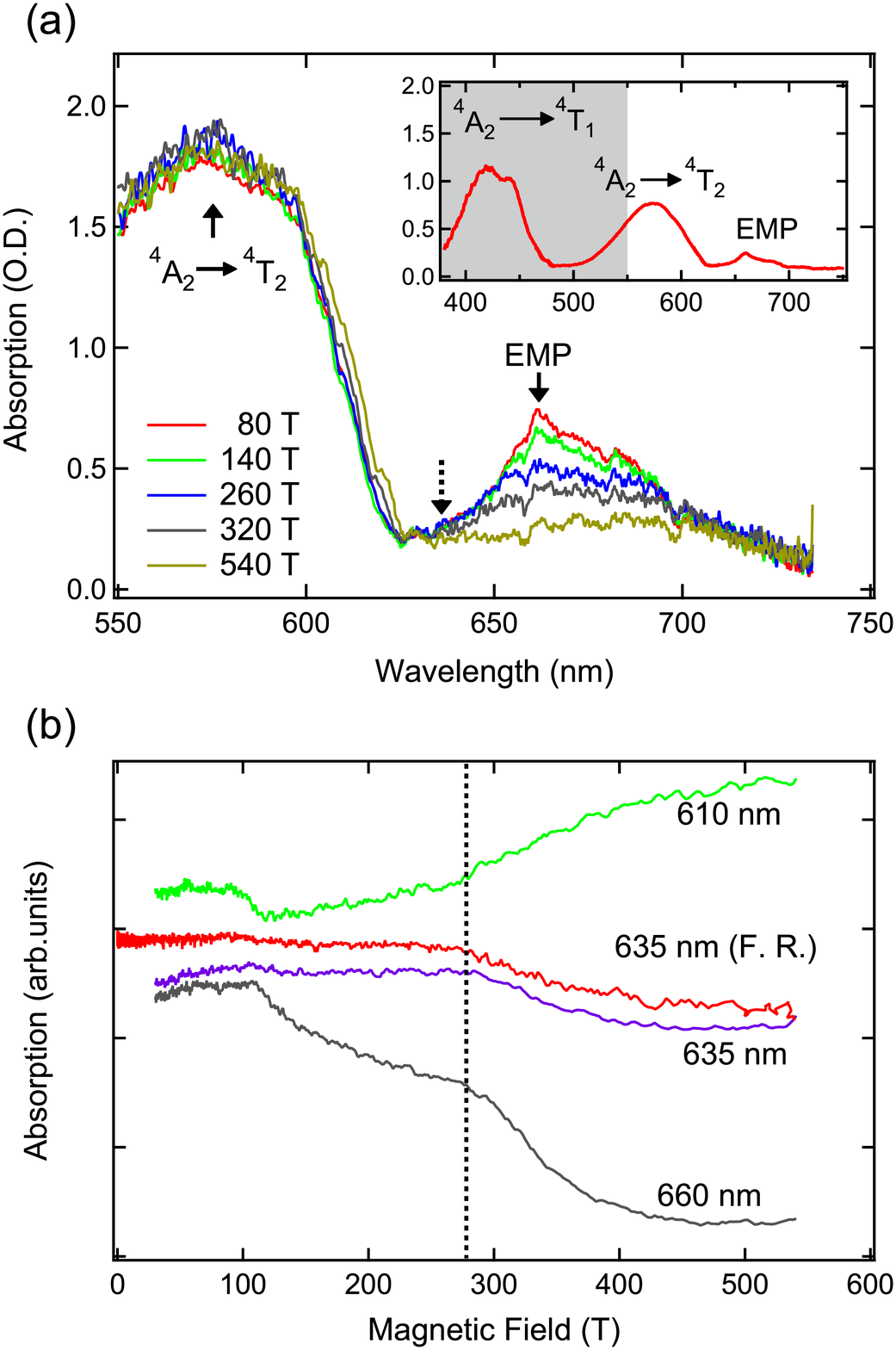}% Here is how to import EPS art
\caption{\label{fig:spectra}  (a) Magneto-optical spectra obtained by streak spectroscopy at 12 K and in several magnetic fields. (b) Absorption intensity changes in several wavelengths at 12 K as a function of magnetic field derived from Fig.~\ref{fig:spectra} (a). The dotted arrow is the point of a wavelength (635 nm) used for the Faraday rotation.
}
\end{figure}

Figure~\ref{fig:spectra} (a) shows the absorption spectra arising from a d-d transition ($^4A_2 \rightarrow ^4T_2$) and an EMP transition in 
Cr$^{3+}$ in several magnetic fields 
obtained from streak spectroscopy. The inset in Fig.~\ref{fig:spectra} (a) presents absorption spectra taken at 0 T with a wider wavelength region 
covering three absorption peaks ($^4A_2 \rightarrow ^4T_2$ transition, $^4A_2 \rightarrow ^4T_1$ transition and an EMP transition in Cr$^{3+}$).  
Figure~\ref{fig:spectra} (b) presents the absorption intensity change in magnetic fields at several chosen wavelengths taken from 
Fig.~\ref{fig:spectra} (a) and also from the absorption obtained in the Faraday rotation measurement. The absorption obtained by the streak 
spectroscopy at a wavelength of 635 nm and that obtained in the Faraday rotation method ($\lambda =635$ nm, dotted arrow shown in 
Fig.~\ref{fig:spectra} (a)) show good agreement. We observed an almost complete suppression of the absorption peak arising from an EMP transition 
when the magnetic field increased up to a region of the fully-polarized ferromagnetic phase. A d-d transition with $\Delta S^z=-1$ 
(exciton, $^4A_2 \rightarrow ^2E, ^2T_1$), and magnon excitation with $\Delta S^z=+1$ were simultaneously excited to conserve the spin momentum in 
the process of EMP transition. However, the magnon with $\Delta S^z=+1$ could not be excited in the fully-polarized ferromagnetic phase, and 
therefore the EMP transition disappeared, which explains the rapid decrease of absorption observed in Fig.~\ref{fig:temp_mag} (b).

Moreover, we observed an anomalous increase of the intensity in the magneto-optical absorption measured at the tail of the $^4A_2 \rightarrow ^4T_2$  
transition synchronized with a decrease of the intensity of the exciton-magnon-phonon transition at 270 T (the dotted line shown in Fig.~\ref{fig:spectra} (b)). This suggests some changes in the crystal structure at 270 T. Previous theoretical~\cite{bergman06} 
and experimental~\cite{matsuda07, matsuda10} works reported that the space group symmetry of the 1/2 plateau state in chromium spinel oxides is 
universally $P4_332$ symmetry, and that the canted 3:1 phase should be at least as low as the $P4_332$ symmetry. $P4_332$ symmetry could be restored 
to $Fd\overline{3}m$ 
symmetry before or at saturation field, since a fully polarized ferromagnetic state has an ideal pyrochlore symmetry 
$Fd\overline{3}m$~\cite{bergman06}. Our results at 270 T 
shown in Fig.~\ref{fig:spectra} (b) present the possibility of a transition into a partially-polarized ferromagnetic state with higher space group 
symmetry than $P4_332$ symmetry. 
S. Kimura {\it et al.} have pointed out by high magnetic field ESR study the possibility of a novel phase between the canted 3:1 phase and the 
ferromagnetic phase in HgCr$_2$O$_4$, where symmetry was higher than $P4_332$ symmetry~\cite{kimura11}. Furthermore, V. Tsurkan {\it et al.} reported ultrasound studies on a similar compound, ZnCr$_2$S$_4$ in high magnetic fields, and insisted that this material should show subsequent lattice transformations as a bond frustrated magnet, despite the absence of any anomalies observed in the magnetization curves~\cite{tsurkan11}. In their results, reflecting the importance of the spin-lattice coupling, the structural phase transition from the tetragonal to cubic phase with a partially-polarized ferromagnetic phase took place at the vicinity of the saturation field, and then the magnetic transition to the fully-polarized ferromagnetic phase occurred at the saturation 
field. Their results are similar to ours at the vicinity of the saturation field. 
These findings imply that a transition from a canted 3:1 state to a partially-polarized ferromagnetic state with higher crystal symmetry than 
$P4_332$ occurs at a magnetic field of 350 T, and then at 410 T a magnetic phase transition results in the fully-polarized ferromagnetic phase as 
shown in Fig.~\ref{fig:magnetization4k}. This novel magnetic phase has never been predicted by existing theories. Higher-order exchange interactions 
among the distant sublattices, and/or more complex spin-lattice interactions could be responsible for the appearance of the novel phase.

We have discussed a physical analogy of the magnetic phase diagram of ZnCr$_2$O$_4$ to that in $^4$He proposed by H. Matsuda and T. Tsuneto~\cite{matsuda70} and K.-S. Liu and M. E. Fisher~\cite{liu73} in Ref.~\cite{miyata11}. The canted 3:1 phase observed in the case of ZnCr$_2$O$_4$ corresponds to a supersolid state and the ferromagnetic (or paramagnetic) phase to a liquid state. The supersolid (intermediate) state could only exist between the solid and the superfluid state in the phase diagram~\cite{matsuda70, liu73}.  Thus, the partially-polarized ferromagnetic phase between 350 T and 410 T is naturally inferred as a superfluid state from the analogy described above.  A direct transition from a ferromagnetic (i.e., liquid) phase to a canted 3:1 (i.e., supersolid) phase requires a simultaneous breaking of the translational and the spin rotational symmetry, which could take place with less probability.

\begin{figure}
   \includegraphics[width=.7\columnwidth]{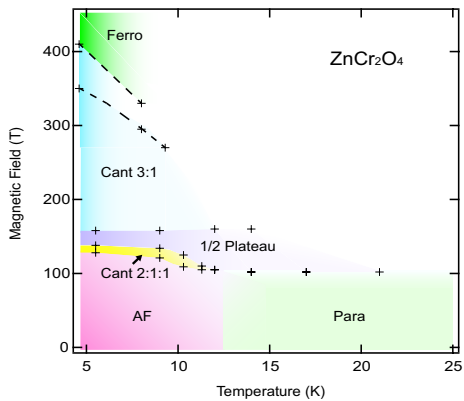}% Here is how to import EPS art
\caption{\label{fig:souzu}Magnetic phase diagram constructed from data in Fig.~\ref{fig:temp_mag} (a) and Ref.~\cite{miyata11}. Cross symbols indicate transition points observed by the Faraday rotation method. Broken lines show anomalies appeared in the magneto-optical absorption.
 }
\end{figure}

Figure ~\ref{fig:souzu} shows a field-temperature phase diagram constructed from the data in Fig.~\ref{fig:temp_mag} and the data obtained by the STC method in our previous paper~\cite{miyata11}. The cross symbols indicate transition points observed by the Faraday rotation method and the broken line shows anomalies observed in the magneto-optical absorption by the Faraday rotation method. Between the broken lines, there should exist a novel magnetic phase. As the temperature increases, a rapid decrease is observed in the magnetic field at which the phase transition takes place between the canted 3:1 phase and the ferromagnetic phase. A similar behavior is observed in other sister materials, CdCr$_2$O$_4$ and HgCr$_2$O$_4$, but the decrease is most pronounced in ZnCr$_2$O$_4$. Thermal fluctuation favors the collinear magnetic structure in the 1/2 plateau phase, or the so-called entropy-driven ``order-by-disorder'' effect~\cite{kawamura84, zhitomirsky02}. This effect becomes pronounced in the small limit of the spin-lattice coupling, and is discussed in Ref.~\cite{miyata11}. The same discussion applies equally to the present case for the ferromagnetic phase.

The Faraday rotation and magneto-optical absorption spectra measurements conducted in an ultra-high magnetic field up to 600 T revealed the full-magnetization processes of ZnCr$_2$O$_4$, and a magnetic phase diagram was obtained. We found anomalies in the absorption spectra as a function of the magnetic field, which implied the certain existence of a novel phase between a canted 3:1 phase and the ferromagnetic phase, corresponding to the superfluid state in the analogy of $^4$He. This phase should have higher symmetry than the $P4_332$ one with a partially-polarized ferromagnetic structure, and will require a more sophisticated model to interpret.
Magneto-absorption was found to be a good tool to detect magnetic phase transitions which are insensitive to the magnetization data.

\end{document}